\documentclass[aps,showpacs,nofootinbib,superscriptaddress,twocolumn]{revtex4}
\usepackage{graphicx,subfigure}
\usepackage{amssymb}
\usepackage{bm}
\usepackage{array}
\usepackage{slashed}
\usepackage{braket}
\usepackage{multirow}
\usepackage{bigstrut}
\usepackage[nooneline,footnotesize,center]{caption2}
\usepackage{longtable}
\usepackage{hyperref}

\begin{document}

\title{Strange quark-antiquark asymmetry of nucleon sea from $\Lambda/\bar\Lambda$ polarization}

\author{Xiaozhen Du}
\affiliation{School of Physics and State Key Laboratory of Nuclear Physics and
Technology, Peking University, Beijing 100871,China}

\author{Bo-Qiang Ma}
\email{mabq@pku.edu.cn}
\affiliation{School of Physics and State Key Laboratory of Nuclear Physics and
Technology, Peking University, Beijing 100871,China}
\affiliation{Collaborative Innovation Center of Quantum Matter, Beijing,China}
\affiliation{Center for High Energy Physics, Peking University, Beijing 100871,China}

\date{\today}

\begin{abstract}
Distinctions between quark to $\Lambda$ and $\bar\Lambda$ longitudinal spin transfers in the semi-inclusive deep inelastic scattering process were observed by the E665 and COMPASS Collaborations. There are suggestions that the difference between $\Lambda$ and $\bar\Lambda$ production is related to the asymmetric strange-antistrange distribution inside the nucleon. However, previous calculations are still too small to explain the experimental data. From a realistic consideration of quark to $\Lambda$ fragmentation due to different flavors, we investigate the strange quark contribution for $\Lambda$ production and polarization. We find that the strange quark-antiquark asymmetry of the nucleon sea can be amplified into an observable quantity from the difference between $\Lambda$ and $\bar\Lambda$ polarizations after taking into account the larger probability of the $\Lambda$ produced from the $s$ quark fragmentation process compared to that from the $u$ or $d$ quark. The qualitative agreement between our calculation and the experimental data supports the existence of the intrinsic strange sea and the strange-antistrange asymmetry. Thus the polarization of $\Lambda/\bar\Lambda$ does open a new window to probe the nucleon sea properties, especially the strange content and its quark-antiquark asymmetry.
\end{abstract}



\maketitle


The sea content of nucleons is a key focus in hadronic physics as there exist a number of anomalies in experiments, such as the proton spin problem~\cite{Ashman:1989ig} and the NuTeV anomaly~\cite{Zeller:2001hh,Zeller:2002du}. Due to the nonperturbative nature of quantum chromodynamics(QCD) at low energy scale, it is complicated to theoretically calculate the sea properties and structures. Also, the limited data sensitive to the sea content of the nucleon from experiments makes the detailed sea structure remain obscure. Even the existence of the intrinsic sea quark and the strange quark-antiquark asymmetry are still under controversy. It is thus important to find experimental quantities that are sensitive to revealing the striking features of the nucleon sea content.

From theoretical aspects, the existence of the intrinsic charm quark and the non-negligible $uudc\bar c$ Fock component in the proton was proposed in 1980 to explain the unexpectedly large cross section of charmed hadrons at high $x_F$~\cite{Brodsky:1980pb}. This intrinsic quark model has been extended to the light-quark sector with significant progress recently~\cite{Chang:2011vx}. Under the frame of the intrinsic quark sea, several theoretical models were proposed to investigate the nonperturbative contributions to the nucleon asymmetric strange-antistrange sea distribution, such as the baryon-meson fluctuation model, the meson cloud model, and the chiral quark model~\cite{Brodsky:1996hc,Signal:1987gz,Burkardt:1991di,Holtmann:1996be,Ma:1997gh,Christiansen:1998dz}. This asymmetric distribution was thought to play an important role in the extraction of the Weinberg angle~\cite{Davidson:2001ji} from neutrino-nucleon deep inelastic scattering , and it has the potential to remove the NuTeV anomaly~\cite{Abbaneo:2001ix,Kretzer:2003wy,Ding:2004ht,Alwall:2004rd,Ding:2004dv,Wakamatsu:2004pd,Ball:2009mk,Ball:2011mu}.

From experimental aspects, the current constraints on the strangeness mainly come from the neutrino dimuon production data. This is because a neutrino can resolve the flavor of the nucleon constituents and such an ability can be used to isolate the nucleon strange-antistrange quark distributions. CCFR and NuTeV dimuon measurements~\cite{Bazarko:1994tt,Rabinowitz:1993xx,Mason:2004yf,Olness:2003wz,Mason:2007zz,KayisTopaksu:2002xq} are such experiments. However, these experiments are not enough to provide a meaningful fit for the strangeness distributions since there are large uncertainties~\cite{Gao:2005gj}. The semi-inclusive deep inelastic scattering~(SIDIS) process also has the power to probe the strangeness in the nucleon. The HERMES Collaboration has recently reported the determination of $x(s(x) + \bar s(x))$ over the range of $0.02 < x < 0.5$ at $Q^2 = 2.5~\mathrm {GeV}^2$ from the measurement of charged kaon production on a deuteron target~\cite{Airapetian:2013zaw}.
Apart from the kaon production, the $\Lambda/\bar\Lambda$ hyperon produced in high energy interactions can also supply information concerning the strange content of the nucleon~\cite{Lu:1995np,Ellis:1995fc,Ma:2000uv,Anselmino:2001ps,Chi:2014xba}.
This is due to the fact that $\Lambda$ is the lightest baryon containing a strange quark; it has a relatively large production cross section, and its polarization can be also determined by measuring the angular distributions of the decay products.

The measurement of the polarization of $\Lambda/\bar\Lambda$ hyperons not only allows a unique test of the quark distributions of the target nucleon but also the quark to $\Lambda/\bar\Lambda$ fragmentation functions~\cite{Jaffe:1996wp,Kotzinian:1997vd,deFlorian:1997zj,Boros:1998kc,Ma:1998pd,Ma:1999wp,Ma:2000uu,Anselmino:2000ga,Anselmino:2000vs,Anselmino:2001ey,Boros:2000ex,Ellis:2007ig,Zhou:2009mx,Chi:2013hka}.
More explicitly, the studies of $\Lambda/\bar\Lambda$ polarization can provide us a lot of useful information, such as the quark to $\Lambda/\bar\Lambda$ fragmentation functions~\cite{Kotzinian:1997vd,deFlorian:1997zj,Ma:1998pd,Anselmino:2000ga,Anselmino:2000vs,Anselmino:2001ey,Boros:2000ex,Chi:2013hka} and the quark distribution functions in nucleon~\cite{Ellis:1995fc,Jaffe:1996wp,Boros:1998kc,Ma:1999wp,Ma:2000uu}, especially in regard to the strange densities~\cite{Lu:1995np,Ma:2000uv,Anselmino:2001ps,Ellis:2007ig,Chi:2014xba,Zhou:2009mx}.
For example, it is suggested that the difference between $\Lambda$ and $\bar\Lambda$  production is related to the asymmetric strange-antistrange
distribution inside the nucleon~\cite{Ma:2000uv,Anselmino:2001ps}. There have been some calculations trying to reveal the different behaviors of $\Lambda$ and $\bar\Lambda$ from the asymmetric strange-antistrange distribution~\cite{Ma:2000uv,Zhou:2009mx,Chi:2014xba}. However, results from Refs.~\cite{Ma:2000uv} and \cite{Chi:2014xba} are still too small to explain the experimental data. Reference~\cite{Zhou:2009mx} produces a large difference between $\Lambda$ and $\bar\Lambda$ polarization, with results close to the existing COMPASS data; while in their calculation the extreme case for $S^-(x)=s(x)-\bar s(x)$, i.e., $S^-(x)=s(x)+\bar s(x)$ or $S^-(x)=-(s(x)+\bar s(x))$ was adopted.

In this paper, we present a general analysis of the spin transfer for the $\Lambda$ and $\bar\Lambda$ hyperons both in $e^+e^-$ annihilation and SIDIS processes. We point out that the small magnitude of the spin transfer compared to the E665~\cite{Adams:1999px} and the COMPASS~\cite{Alekseev:2009ab} data in previous studies~\cite{Ma:2000uv,Chi:2014xba} is due to the equivalent treatment of the fragmentations of $s,~u$ and $d$ quarks by using the Gribov-Lipatov relation~\cite{Barone:2000tx,Gribov:1971zn}. However, the probability of the $\Lambda$ produced from the $s$ quark fragmentation process should be larger than that from the $u$ or $d$ quark~\cite{deFlorian:1997zj}. This effect can be taken into account by relating the model result of quark fragmentations to the Albino, Kniehl, and Kramer~(AKK) parametrization~\cite{Albino:2008fy}. To verify the reasonability of our consideration of the strange quark fragmentation process, we first calculate the $\Lambda$ polarization in $e^+e^-$ annihilation at the Z-pole, then we extend our calculation to the SIDIS process at COMPASS, E665, and HERMES.
Although the results we get from a realistic consideration of the contributions by different flavor quarks based on the quark-spectator-diquark model have no quantitative value, the qualitative agreement between our calculations and the experimental data suggests an amplified contribution from strange quarks, so that the strange quark-antiquark asymmetry of the nucleon sea can produce an observable difference between $\Lambda$ and $\bar{\Lambda}$ polarizations observed in experimental data, as was expected from theoretical considerations~\cite{Ma:2000uv,Anselmino:2001ps}. Thus the $\Lambda$ and $\bar{\Lambda}$ production in the SIDIS process is indeed an ideal place to probe the nucleon strange sea properties and its quark-antiquark asymmetry. Further measurements of $\Lambda/\bar{\Lambda}$ polarization with accurate enough data should provide us more information on the nucleon strange sea content and give us better control of the form of the quark to $\Lambda/\bar{\Lambda}$ fragmentation functions.

In the $e^+e^-$ annihilation process around the bosonic Z-pole, the $q\bar q$ pair produced from the weak interaction can be polarized even though the initial $e^+e^-$ states are unpolarized. These polarized quark pairs lead to the polarized $\Lambda$ hyperon eventually. Thus the measurement on the polarization of $\Lambda$ can be used to test the quark fragmentations. The expression of the final $\Lambda$ polarization in this process can be given as
\begin{equation}\label{transfer_ani}
P_{\Lambda}=-\frac{\sum_{q}A_q[\Delta D^{\Lambda}_q(z)-\Delta D^{\Lambda}_{\bar q}(z)]}{\sum_{q}C_q[D^{\Lambda}_q(z)+D^{\Lambda}_{\bar q}(z)]},
\end{equation}
where $\Delta D^{\Lambda}_{q(\bar q)}(z)$ and $D^{\Lambda}_{q(\bar q)}(z)$ are the polarized and unpolarized fragmentation functions for a quark $q$ splitting into a $\Lambda$ hyperon with the longitudinal momentum fraction $z$, and the parameters $A_q,~C_q$ can be referred to in Ref.~\cite{Ma:1999wp}.

The final $\Lambda$ hyperon can be produced either from the direct quark fragmentation process or via the decay of heavier resonances, such as $\Sigma^*$,~$\Sigma^0$, and $\Xi$, which can also partially transfer their polarization to the $\Lambda$. Taking both of the possibilities into account, we can rewrite the polarized fragmentation function $\Delta D^{\Lambda}_q(z)$ and the unpolarized fragmentation function $D^{\Lambda}_q(z)$ as
\begin{eqnarray}\label{FF}
\Delta D^{\Lambda}_{q}(z,Q^2)&=&a_1\Delta D_{q}^{\Lambda(\mathrm{direct})}(z,Q^2)+a_2\Delta D_{q}^{\Sigma^{0}}(z^\prime,Q^2)\alpha_{\Sigma^{0}\Lambda}\nonumber\\
&+&a_3\Delta D_{q}^{\Sigma^{\ast}}(z^\prime,Q^2)\alpha_{\Sigma^{\ast}\Lambda}
+a_4\Delta D_{q}^{\Xi}(z^\prime,Q^2)\alpha_{\Xi\Lambda},\nonumber\\
D^{\Lambda}_{q}(z,Q^2)&=&a_1D_{q}^{\Lambda(\mathrm{direct})}(z,Q^2)+a_2D_{q}^{\Sigma^{0}}(z^\prime,Q^2) \nonumber\\
&+&a_3D_{q}^{\Sigma^{\ast}}(z^\prime,Q^2)+a_4D_{q}^{\Xi}(z^\prime,Q^2).
\end{eqnarray}
Here, $a$'s are weight coefficients representing the ratio of $\Lambda$ produced from different channels. In Ref.~\cite{Airapetian:2006ee}, the contribution of the indirectly produced $\Lambda$ was estimated by a Monte Carlo simulation to be as large as 60\%. This is in accord with the calculation using the LEPTO generator, which indicates that only about 40\%-50\% of $\Lambda$ are produced directly, 30\%-40\% are originated from $\Sigma^{\ast}$(1385) decaying, and about 20\% are decay products of the $\Sigma^{0}$~\cite{Adolph:2013dhv}. We choose $a_1=0.4,~a_2=0.2,~a_3=0.3,~a_4=0.1$ in our calculations~\cite{Chi:2014xba}. To see the dependence of our results on the chosen values of the $a$'s parameters, we let $a_1$ have a 10\% variable range with the other $a$'s parameters being adjusted according to its proportion among $a_2,~a_3$, and $a_4$; i.e., $a_1=0.4\pm0.1,~a_2=0.2\mp0.0333,~a_3=0.3\mp0.05,~a_4=0.1\mp0.0167$, and show the results with bands. The decay parameters $\alpha$'s are $\alpha_{\Sigma^{0}\Lambda}=-0.333,~\alpha_{\Sigma^{\ast}(\frac{3}{2},\frac{3}{2})\Lambda}=1.0,~\alpha_{\Sigma^{\ast}(\frac{3}{2},\frac{1}{2})\Lambda}=0.333,~\alpha_{\Xi^{0}\Lambda}=-0.406,~\alpha_{\Xi^{-}\Lambda}=-0.458$, and $z^\prime=1.1z$, based on the same consideration with those of Ref.~\cite{Chi:2014xba}.

For the fragmentation functions of hyperons appearing in Eq.~(\ref{FF}), they cannot be calculated using perturbative QCD directly since fragmentations are nonperturbative processes. At present, there are still no parametrizations about the fragmentations of $\Sigma^*$,~$\Sigma^0$, and~$\Xi$. The AKK collaboration gives fits of $\Lambda/\bar\Lambda$ fragmentation functions, but it does not include the polarized ones~\cite{Albino:2008fy}. In this case, phenomenological models are quite useful in particular in obtaining some guide for experiments. We know that quark fragmentation functions $D_q^h(z)$ and distribution functions $q_h(z)$, which present the probability of finding the corresponding quark $q$ carrying a momentum fraction $z$ inside the same hadron, can be related by the phenomenological Gribove-Lipatov relation~\cite{Barone:2000tx,Gribov:1971zn}
\begin{equation}
D_q^h(z)\sim zq_h(z).
\end{equation}
Although this relation is known to be valid only at a specific $Q^2$ near $z\rightarrow1$, it provides us reasonable guidance for a phenomenological parametrization of the hadron fragmentation functions. In this work, the valence quark distribution functions are analyzed in a SU(6) quark-spectator diquark model, and the sea quark distributions are obtained from the SU(3) symmetry relations between octet baryons by using the CT14 parametrization (CT14 LO)~\cite{Dulat:2015mca}. Since the valence quark distributions in the SU(6) quark-spectator diquark model have been discussed in detail in Ref.~\cite{Chi:2014xba}, we omit the descriptions about this model here. The specific forms of the quark distributions can be referred to in Ref.~\cite{Chi:2014xba}.

As mentioned above, the probability of the $\Lambda$ produced from the $s$ quark fragmentation should be larger than that from a $u$ or $d$ quark. This is because if the $\Lambda$ originates from the primarily $u$ quark, a $s\bar s$ and a $d\bar d$ pair have to be created in order to provide the constituent quarks which should be suppressed with respect to the creation of only $u\bar u$ and $d\bar d$ pairs required if the $\Lambda$ is produced from an initial $s$ quark. However, in the previous study~\cite{Chi:2014xba}, this effect was suppressed, since every flavor quark fragmentation function was normalized to the same form. The ratio of the probability of $\Lambda$ produced from the $s$ quark to that from the $u$ quark is about $0.9$ in Ref.~\cite{Chi:2014xba}, while from the AKK parametrization this ratio is about $1.4$~\cite{Albino:2008fy}. To show this effect, we take the following equations to relate the model quark fragmentations to the AKK parametrization:
\begin{eqnarray}\label{FFadjust}
D_{u}^{\Lambda}(x,Q^2) &=& D_{d}^{\Lambda}(x,Q^2)\nonumber\\
&=&\left (\frac{D_{u}^{\Lambda}(x)}{D_{u+\bar u}^{\Lambda}(x)}\right)^{\mathrm{th}}D_{u+\bar u}^{\Lambda}(x,Q^2)^{\mathrm{AKK}},\nonumber\\
D_{\bar u}^{\Lambda}(x,Q^2) &=& D_{\bar d}^{\Lambda}(x,Q^2)\nonumber\\
&=&\left (\frac{D_{\bar u}^{\Lambda}(x)}{D_{u+\bar u}^{\Lambda}(x)}\right)^{\mathrm{th}}D_{u+\bar u}^{\Lambda}(x,Q^2)^{\mathrm{AKK}},\nonumber\\
\Delta D_{u}^{\Lambda}(x,Q^2)&=&\Delta D_{d}^{\Lambda}(x,Q^2)\nonumber\\
&=& \left(\frac{\Delta D_{u}^{\Lambda}(x)}{D_{u+\bar u}^{\Lambda}(x)}\right)^{\mathrm{th}}D_{u+\bar u}^{\Lambda}(x,Q^2)^{\mathrm{AKK}},\nonumber\\
D_{s}^{\Lambda}(x,Q^2)&=& \left(\frac{D_{s}^{\Lambda}(x)}{D_{s+\bar s}^{\Lambda}(x)}\right)^{\mathrm{th}}D_{s+\bar s}^{\Lambda}(x,Q^2)^{\mathrm{AKK}},\nonumber\\
D_{\bar s}^{\Lambda}(x,Q^2)&=& \left(\frac{D_{\bar s}^{\Lambda}(x)}{D_{s+\bar s}^{\Lambda}(x)}\right)^{\mathrm{th}}D_{s+\bar s}^{\Lambda}(x,Q^2)^{\mathrm{AKK}},\nonumber\\
\Delta D_s^{\Lambda}(x,Q^2) &=& \left(\frac{\Delta D_s^{\Lambda}(x)}{D_{s+\bar s}^{\Lambda}(x)}\right)^{\mathrm{th}}D_{s+\bar s}^{\Lambda}(x,Q^2)^{\mathrm{AKK}}.
\end{eqnarray}
This approach is similar to the way we adopted the quark distribution functions in Ref.~\cite{Chi:2014xba}.

\begin{figure}
\centering
\includegraphics[width=1.0\columnwidth]{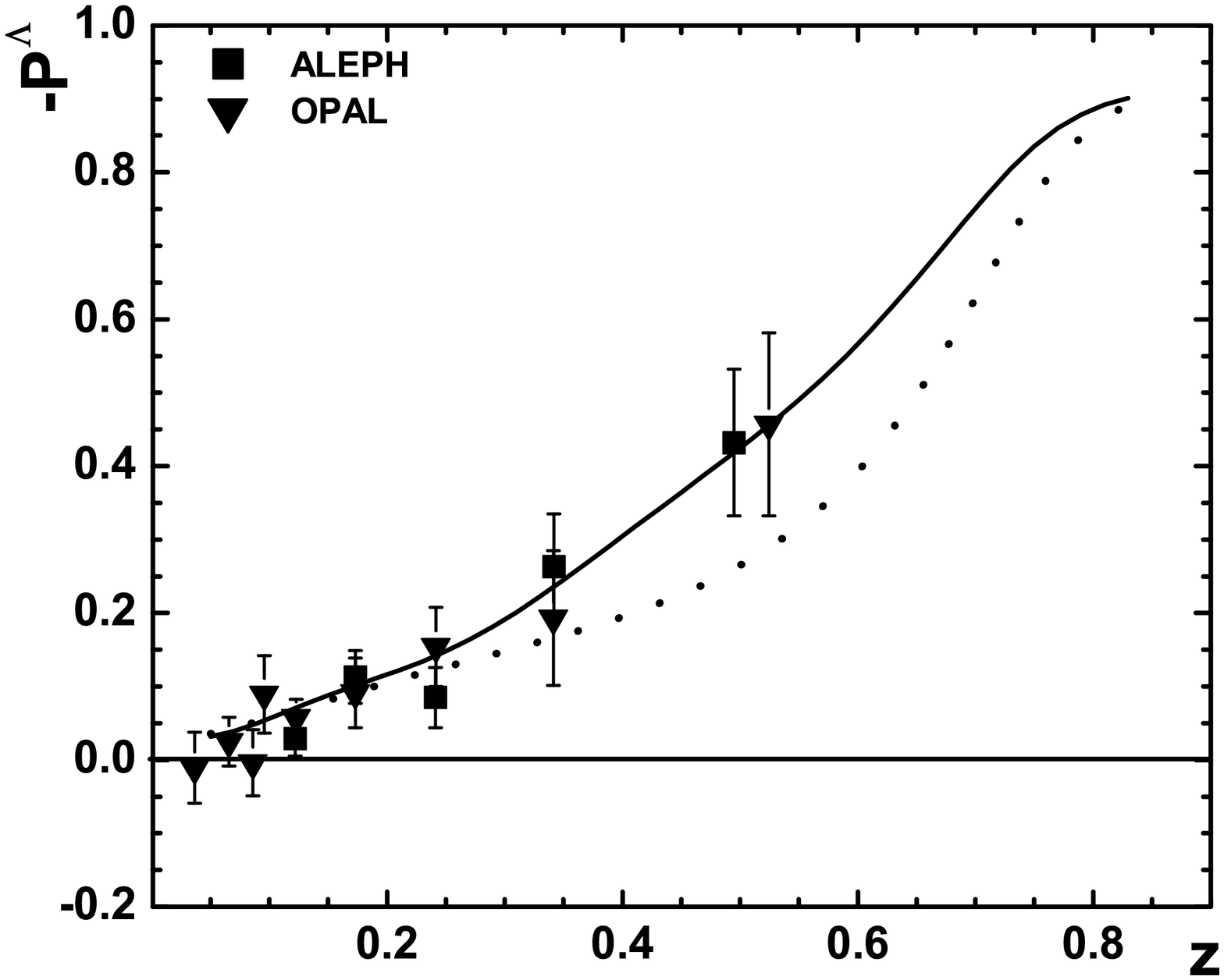}
\caption{\baselineskip 13pt The $z$ dependence of $\Lambda$ polarization in $e^+e^-$ annihilation process. The solid curve corresponds to the theoretical calculation with AKK parametrization input, and the dotted curve is from the previous work without AKK parametrization input~\cite{Chi:2013hka}.}
\label{fig:lep} 
\end{figure}

The solid line in Fig.~\ref{fig:lep} shows the results of $z$ dependence of a $\Lambda$ hyperon in $e^+e^-$ annihilation process at the $Z$ pole~\cite{Buskulic:1996vb,Ackerstaff:1997nh}, while the dotted line corresponds to the previous work without the AKK parametrization input~\cite{Chi:2013hka}. To compare with the dotted line, the $a$'s parameters should be taken as the same as those of the previous work~\cite{Chi:2013hka}. Therefore, we do not show the bands in this figure due to the 10\% variable ranges of the $a_1$ parameter. It is interesting to notice that, compared to the dotted line, the solid line with the AKK parametrization input has a much better description of the experimental data, especially in the medium $z$ region $0.2 \to 0.6$. This suggests that the disagreement between the data and the dotted line in the previous study could be partially explained by the larger probability of the $s$ quark fragmentation.

Now we extend our calculation to the SIDIS process. For a longitudinal polarized charged lepton beam and an unpolarized nucleon target, the longitudinal spin transfer to the $\Lambda$ hyperon is
\begin{eqnarray}\label{transfer_dis}
A^{\Lambda}(z)=\frac{\int\mathrm{d}x\mathrm{d}y\frac{S x}{Q^4}
\sum_{q}e_q^2f_q(x,Q^2)\Delta D^{\Lambda}_q(z,Q^2)}
{\int\mathrm{d}x\mathrm{d}y\frac{S x}{Q^4}\sum_{q}e_q^2f_q(x,Q^2)D^{\Lambda}_q(z,Q^2)},
\end{eqnarray}
where $e_q$ is the electric charge of parton $q$, $x=Q^2/(2P\cdot q),~y=P\cdot q/(P\cdot \ell)$, and $z=P\cdot P_h/(P\cdot q)$ are three Lorentz invariant variables, and $Q^2=Sxy$ is the squared four-momentum transfer of the virtual photon with $S=M_p^2+m_\ell^2+2M_pE_\ell$. For the $\bar\Lambda$ hyperon, the spin transfer $A^{\bar\Lambda}$ can be obtained from Eq.~(\ref{transfer_dis}) by replacing $q$ with $\bar q$ and $\Lambda$ with $\bar\Lambda$. After integrating the numerator and denominator on $x$ and $y$ respectively, we can get the longitudinal spin transfer of $\Lambda$ or $\bar\Lambda$ as a function of $z$.

At COMPASS, the given measurement cuts are about $x,~y$, and the Feynman variable $x_F$~\cite{Alekseev:2009ab}: $1~\mathrm{GeV}^2<Q^2<50~\mathrm{GeV}^2,~0.005<x<0.65,~0.2<y<0.9,~0.05<x_{_F}<0.5$. To get the spin transfer dependence on $x_F$, we express $z$ as a function of $x,~y$, and $x_F$ with
\begin{equation}\label{zrelation}
z=\frac{x_F}{2}\sqrt{\frac{4M^2x}{Sy}+1}+\left(\frac{M^2}{Sy}+\frac{1}{2}\right)\sqrt{\frac{4(M_h^2+P_{h\perp}^2)}{M^2+Sy-Sxy}+x^2_F},
\end{equation}
where $P_{h\perp}$ is the transversal momentum of the final $\Lambda$ hyperon. We set the value of $P_{h\perp}$ to be $3.0$~GeV which is in a reasonable region compared to the COMPASS experimental center-of-mass energy, $S=320~\mathrm{GeV}^2$.

In our numerical calculation, the quark distribution functions of the $u$ and $d$ flavors in Eq.~(\ref{transfer_dis}) are from the CTEQ parametrization~\cite{Dulat:2015mca}. For strange quarks, the CTEQ parametrization of $s$ and $\bar s$ are flavor blind, and the results are in fact an average of them. To investigate the contribution to the spin transfer difference between $\Lambda$ and $\bar\Lambda$ hyperon from the nucleon asymmetric strange-antistrange sea distribution, we need an asymmetric nucleon strange sea input. We consider the nucleon asymmetric strange-antistrange sea distribution effect in the meson-baryon fluctuation model~\cite{Brodsky:1996hc}, where the nucleon wave function at low resolution can be viewed as a fluctuating system coupling to the intermediate noninteracting baryon-meson Fock states. As is mentioned in Ref.~\cite{Chi:2014xba}, this way of determining $s(x)$ and $\bar s(x)$ does not take the QCD evolution effect into account. However it is pointed out in Refs.~\cite{Moch:2004pa,Catani:2004nc} that the quark-antiquark symmetry of the nucleon sea would be violated when the perturbative QCD evolution was calculated at the next-to-next-to-leading order, where the splitting functions for quarks and antiquarks are different from each other. To reflect the QCD evolution effect while keeping $s(x,Q^2)+\bar s(x,Q^2)$ the same with that of the CTEQ parametrization, a reasonable form of the nucleon strange sea input is given as~\cite{Chi:2014xba}
\begin{eqnarray} \label{newfluctuation}
s^P(x,Q^2)&=&\frac{2s^{\mathrm{th}}(x)}{s^{\mathrm{th}}(x)+\bar s^{\mathrm{th}}(x)}s^{\mathrm{ctq}}(x,Q^2),\nonumber\\
\bar s^P(x,Q^2)&=&\frac{2\bar s^{\mathrm{th}}(x)}{s^{\mathrm{th}}(x)+\bar s^{\mathrm{th}}(x)}s^{\mathrm{ctq}}(x,Q^2),
\end{eqnarray}
where $s^{\mathrm{th}}$ and $\bar s^{\mathrm{th}}$ are the strange and antistrange quark distributions from the baryon-meson fluctuation model~\cite{Brodsky:1996hc}, and $s^{\mathrm{ctq}}(x)$ is the strange quark distribution from the CTEQ parametrization~\cite{Dulat:2015mca}.

\begin{figure}
\centering
\includegraphics[width=1.0\columnwidth]{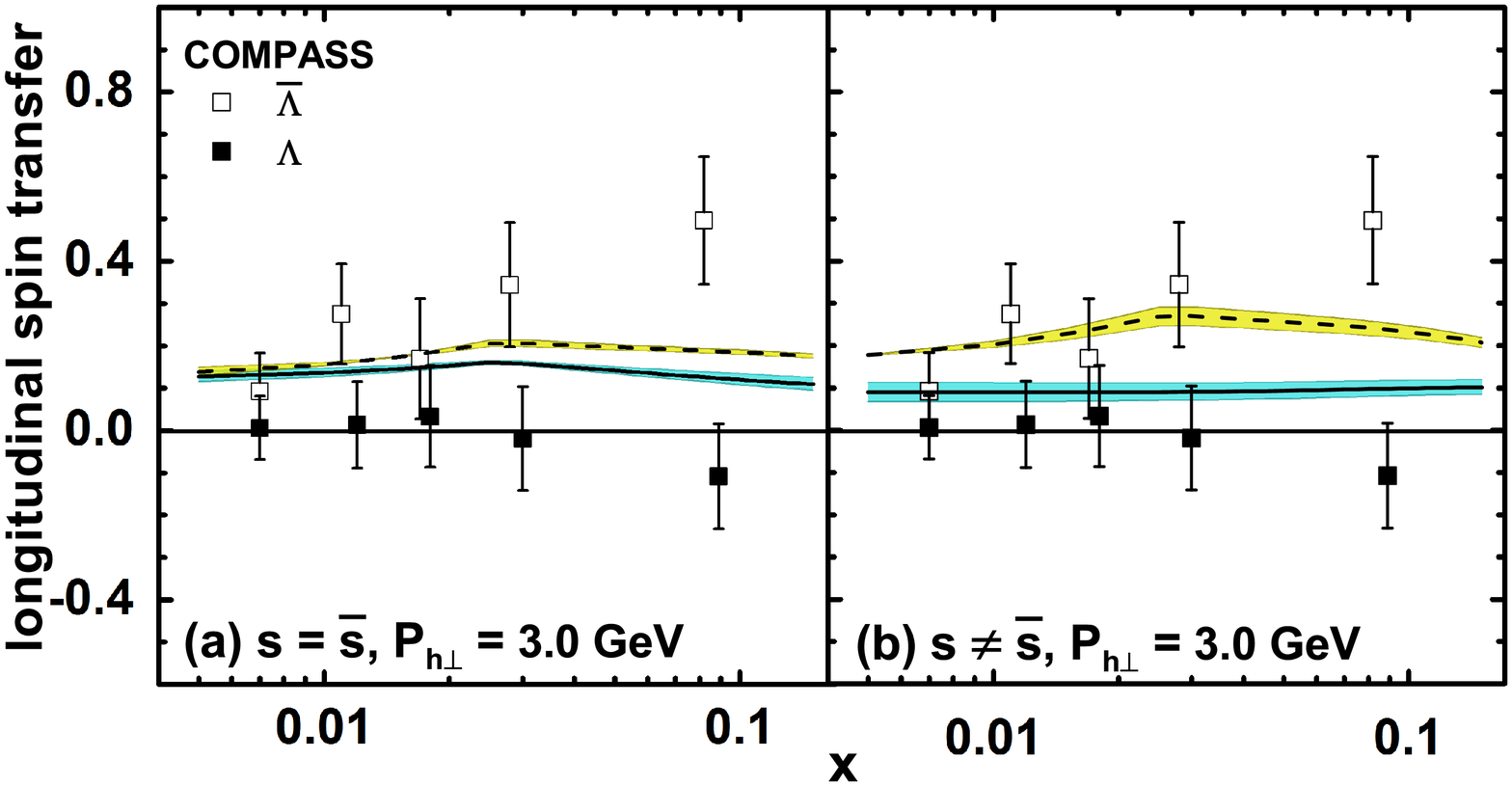}
\caption{\baselineskip 13pt The $x$ dependence of $\Lambda$ and $\bar \Lambda$ longitudinal spin transfers with and without the asymmetric strange-antistrange sea distribution input at COMPASS. The solid and dashed curves correspond to $\Lambda$ and $\bar\Lambda$ hyperons. The bands are from 10\% variable ranges of $a_1$ parameter.}
\label{fig:compass_x} 
\end{figure}
\begin{figure}
\centering
\includegraphics[width=1.0\columnwidth]{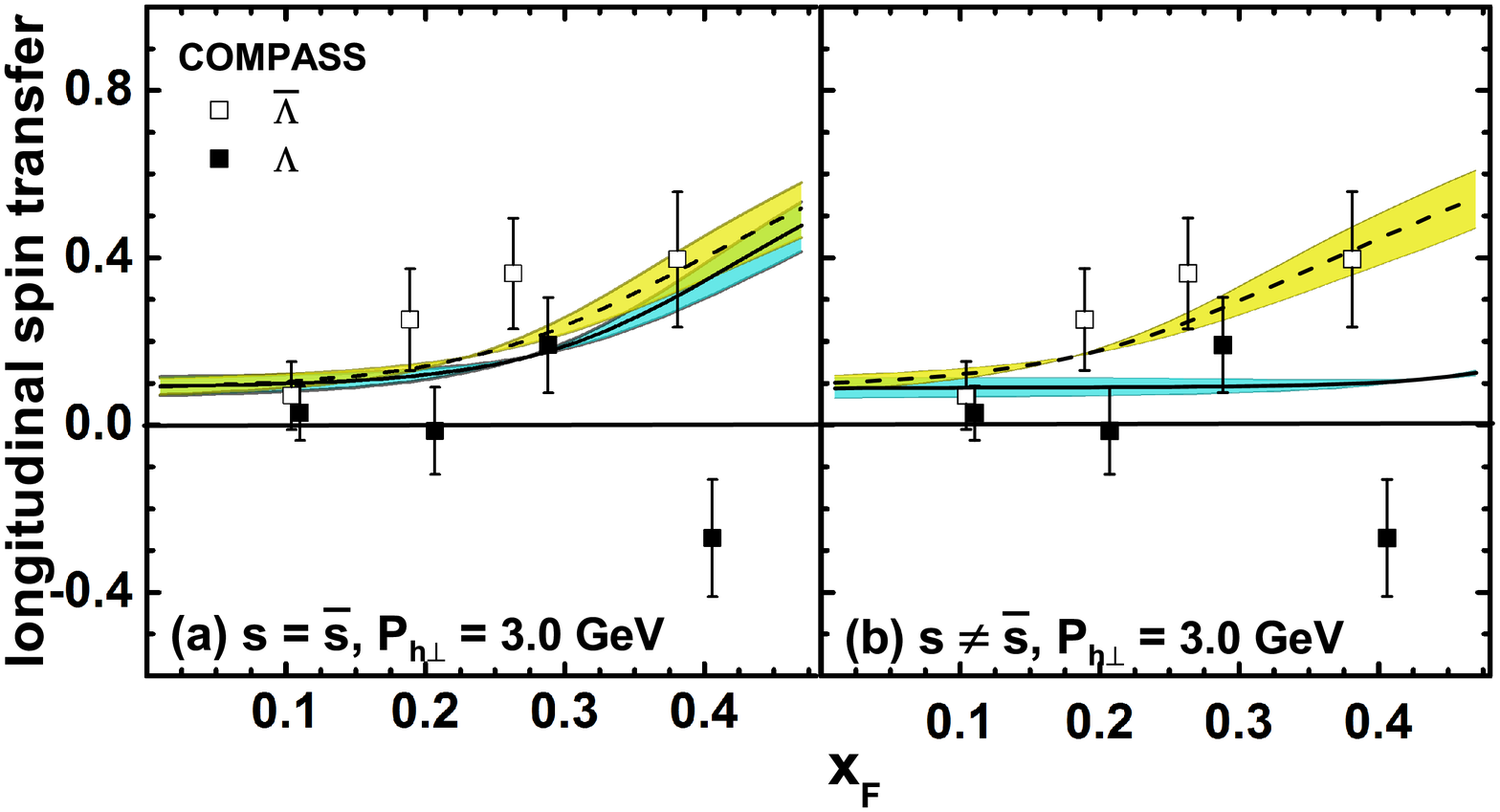}
\caption{\baselineskip 13pt The $x_F$ dependence of $\Lambda$ and $\bar \Lambda$ longitudinal spin transfers with and without the asymmetric strange-antistrange sea distribution input at COMPASS. The solid and dashed curves correspond to $\Lambda$ and $\bar\Lambda$ hyperons. The bands are from 10\% variable ranges of $a_1$ parameter.}
\label{fig:compass_xf} 
\end{figure}

Figures~\ref{fig:compass_x} and \ref{fig:compass_xf} show the results of $x$ and $x_F$ dependences of $\Lambda$ and $\bar\Lambda$ longitudinal spin transfers, together with the COMPASS data. The solid and dashed lines correspond to the $\Lambda$ and $\bar\Lambda$ longitudinal spin transfers, respectively. The bands are from 10\% variable ranges of $a_1$ parameter. In order to focus on the contribution of the nucleon asymmetric strange-antistrange sea distribution, we first calculate the longitudinal spin transfers on the symmetric strange sea input and show the results in the left panels of Figs.~\ref{fig:compass_x} and \ref{fig:compass_xf}. In the right panels of Figs.~\ref{fig:compass_x} and \ref{fig:compass_xf}, the asymmetric strange-antistrange sea distribution is considered. As we can see, the difference between $\Lambda$ and $\bar\Lambda$ gets enlarged under an asymmetric strange-antistrange sea input so that the agreement between our calculation and the data is improved significantly.
Since there are limited data sensitive to the strange sea content, the qualitative agreement between our calculation and the data can be seen as support of the existence of the nucleon strange-antistrange sea distribution. What is more, bands in these two figures tell us that, although the $a$'s values adjusted according to the Monte Carlo predictions~\cite{Adolph:2013dhv} are somewhat arbitrary, the small difference of these weight coefficients does not affect our qualitative prediction and conclusion by the fact that only the shapes of the $\Lambda$ and $\bar{\Lambda}$ longitudinal spin transfers are changed. In fact, we even calculate the $\Lambda$ and $\bar\Lambda$ polarization~(not presented in this work) in the case that there are no intermediate, heavier hyperon decay processes, i.e., $a_1=1.0,~a_2=0,~a_3=0,~a_4=0$, and find that the difference between the quark to $\Lambda$ and the $\bar\Lambda$ longitudinal spin transfers predicted by our model still remains.

\begin{figure}
\centering
\includegraphics[width=1.1\columnwidth]{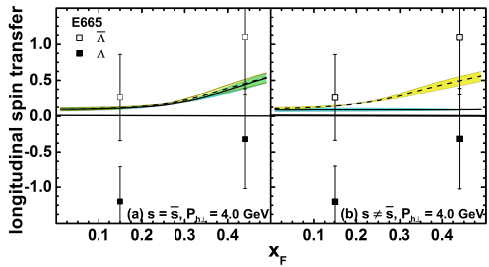}
\caption{\baselineskip 13pt The $x_F$ dependence of $\Lambda$ and $\bar \Lambda$ longitudinal spin transfers with and without the asymmetric strange-antistrange sea distribution input at E665. The solid and dashed curves correspond to $\Lambda$ and $\bar\Lambda$ hyperons. The bands are from 10\% variable ranges of $a_1$ parameter.}
\label{fig:e665_xf} 
\end{figure}

 Next we calculate the $x_F$ dependence of $\Lambda$ and $\bar\Lambda$ longitudinal spin transfers in the E665 experiment. The similar phenomenon with the COMPASS $x_F$ dependent spin transfer appears in our calculations, and is displayed in Fig.~\ref{fig:e665_xf}. The value of $P_{h\perp}$ is set to be $4.0$~GeV, which is also in a reasonable region compared to the E665 experimental center-of-mass energy $S=940~\mathrm{GeV}^2$.

The $x$ and $x_F$ dependences of a $\Lambda$ longitudinal spin transfer have also been measured by the HERMES Collaboration within the kinematical domains~\cite{Airapetian:2006ee,Belostotski:2011zza}: $0.8\mathrm {GeV}^2<Q^2<24\mathrm {GeV}^2$,~$W^2>4\mathrm {GeV}^2$,~$0.05<y<0.9$. We now perform our calculation under the HERMES experimental conditions. Using Eq.~(\ref{zrelation}), we can evaluate the region of $P_{h\perp}$ value from the measured $x,~y,~z$, and $x_F$ bins at HERMES, and the result is about $0.3\to 0.9$~GeV. In our calculation, we choose $P_{h\perp}=0.85$~GeV.
\begin{figure}
\centering
\includegraphics[width=1.0\columnwidth]{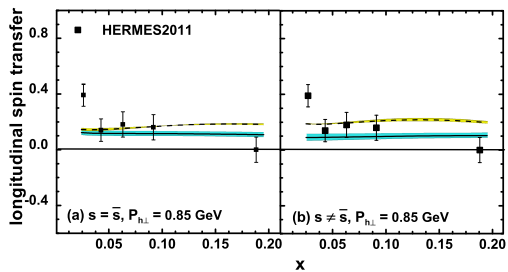}
\caption{\baselineskip 13pt The $x$ dependence of $\Lambda$ and $\bar\Lambda$ longitudinal spin transfers  with and without the asymmetric strange-antistrange sea distribution input at HERMES. The solid and dashed curves correspond to $\Lambda$ and $\bar\Lambda$ hyperons. The bands are from 10\% variable ranges of $a_1$ parameter.}
\label{fig:hermes_x} 
\end{figure}

\begin{figure}
\centering
\includegraphics[width=1.0\columnwidth]{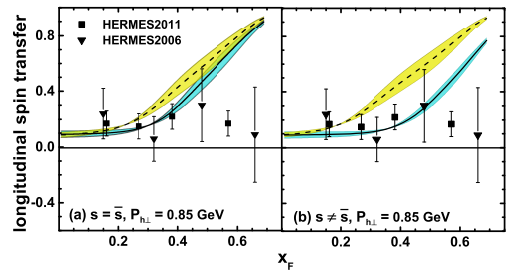}
\caption{\baselineskip 13pt The $x_F$ dependence of $\Lambda$ and $\bar\Lambda$ longitudinal spin transfers  with and without the asymmetric strange-antistrange sea distribution input at HERMES. The solid and dashed curves correspond to $\Lambda$ and $\bar\Lambda$ hyperons. The bands are from 10\% variable ranges of $a_1$ parameter.}
\label{fig:hermes_xf} 
\end{figure}

We show the $x$ and $x_F$ dependence of $\Lambda$ and $\bar\Lambda$ longitudinal spin transfers under the HERMES experimental cuts in Figs.~\ref{fig:hermes_x} and \ref{fig:hermes_xf}. Specially, Figs.~\ref{fig:hermes_x}(a) and \ref{fig:hermes_xf}(a) give the results from the integration without considering the asymmetric nucleon strange sea distribution input, while Figs.~\ref{fig:hermes_x}(b) and \ref{fig:hermes_xf}(b) show the results with this effect taken into account. As expected~\cite{Ma:2000uv,Anselmino:2001ps}, the asymmetric nucleon strange sea input produces a larger spin transfer difference between $\Lambda$ and $\bar\Lambda$ hyperons. Since there are only $\Lambda$ polarization data at HERMES, we also suggest an analysis on the $\bar\Lambda$ polarization to give a more precise examination of the nucleon strange-antistrange sea distribution.

In conclusion, we investigated the $\Lambda$ and $\bar\Lambda$ polarizations in the $e^+e^-$ annihilation process and in lepton-nucleon SIDIS process. The larger probability of a $\Lambda/\bar\Lambda$ hyperon produced from $s/\bar s$ quark fragmentation compared to $u/\bar u$ or $d/\bar d$ flavor was considered. We particularly discuss the importance of the asymmetric nucleon strange-antistrange sea distribution in reproducing the longitudinal spin transfer difference between $\Lambda$ and $\bar\Lambda$ hyperons. We present the results obtained at COMPASS, E665, and HERMES with and without the asymmetric nucleon strange sea distribution input. Compared to Refs.~\cite{Ma:2000uv,Zhou:2009mx,Chi:2014xba}, the asymmetric strange-antistrange sea distribution input and the $s/\bar s$ quark fragmentation process are treated more reasonably in this work. Our results show that the asymmetric nucleon strange sea distribution input gives a better description of the experimental data. This qualitative agreement between the data and our calculation can be viewed as support of the existence of the intrinsic strange sea and the strange-antistrange asymmetry. Since it is difficult to measure the strange content of the nucleon sea, the analysis on the polarization of $\Lambda/\bar\Lambda$ does open a new window to probe the nucleon sea properties. We suggest future experiments to analyze the polarization of $\Lambda/\bar\Lambda$ for providing more information on the nucleon strange sea content.

We would like to thank Tianbo Liu and Jun Zhang for helpful discussions. This work is partially supported by National Natural Science Foundation of China (Grants No.~11120101004 and No.~11475006).


\end{document}